\documentclass[twoside,a4paper,11pt]{proceedings}
\usepackage{graphicx}
\usepackage{hyperref}
\usepackage{movie15}
\usepackage{natbib}
\begin{document}
\pagenumbering{arabic}
\pagestyle{myheadings}
\thispagestyle{empty}
\vspace*{-1cm}
{\flushleft\includegraphics[width=3cm,viewport=0 -30 200 -20]{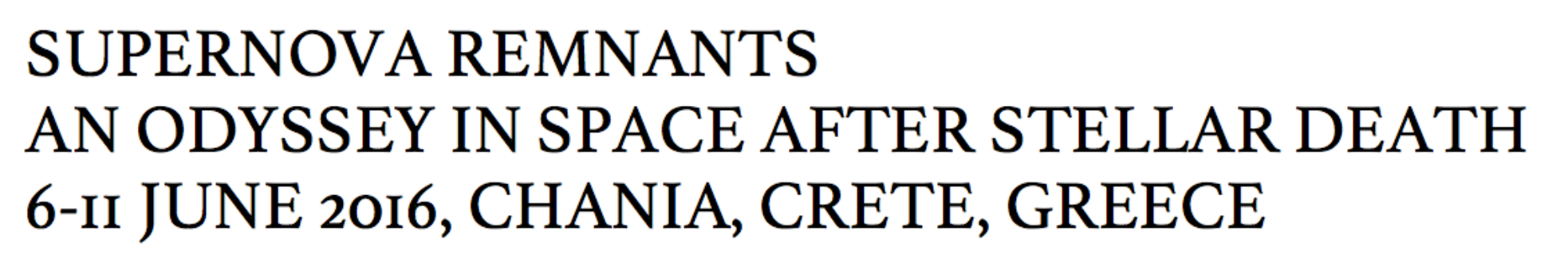}}
\vspace*{0.2cm}
\begin{flushleft}
{\bf {\LARGE
HI absorption spectra for Supernova Remnants in the VGPS survey
}\\
\vspace*{1cm}
D.A. Leahy$^1$
and TS. Ranasinghe$^1$
%
}\\
\vspace*{0.5cm}
%
$^{1}$
University of Calgary, Calgary, Alberta, Canada \\
%
\end{flushleft}
\markboth{
HI absorption spectra for Supernova Remnants
}{
Leahy \& Ranasinghe
}
\thispagestyle{empty}
\vspace*{0.4cm}
\begin{minipage}[l]{0.09\textwidth}
\ 
\end{minipage}
\begin{minipage}[r]{0.9\textwidth}
\vspace{1cm}
\section*{Abstract}{\small
The set of supernova remnants (SNR) from Green's SNR catalog which are found in the VLA Galactic Plane Survey (VGPS) are the objects considered in this study. 
For these SNR, we extract and analyse HI absorption spectra in a uniform way and construct a catalogue of absorption spectra and distance determinations. 
\vspace{10mm}
\normalsize}
\end{minipage}

\section{Introduction}

$\,\!$\indent 

Supernova remnants (SNR) are an important research topic for astrophysics. 
They yield valuable information relevant to stellar evolution, 
the evolution of the Galaxy, and its interstellar medium. 
SNRs are the dominant source of kinetic energy input into the interstellar
medium. 
A SNR emits in X-rays from its hot shocked gas, with temperature of order 1 keV. It emits radio continuum emission from relativistic electrons accelerated
at the SNR shock. 

Finding the distance to a SNR is needed to determine its physical properties, 
such as luminosity, size, and age. 
The distance to a SNR can be estimated using its H I absorption spectrum combined with a model of Galactic rotation.
In this paper we report the initial results from a study of the supernova remnants  in the VLA Galactic Plane Survey (VGPS, Stil et al, 2006).

\section{Data analysis and methods}

$\,\!$\indent The VGPS survey is a 1420 MHz radio continuum survey of the Galactic
plane between longitudes of 18 and 67 degrees, and latitude coverage ranging 
from $-1.3 < b < 1.3$ degrees to $-2.3 < b < 2.3$ degrees.
It also includes HI line images with 0.824 km/s width velocity channels.
The spatial resolution of the survey is 1 arc-minute.

A list of supernova remnants in the VGPS survey area was compiled and the
VGPS data for these SNRs were analysed to obtain HI absorption spectra.
The methods for HI absorption spectrum analysis are described in  Tian, Leahy and Wang (2007) and Leahy and Tian (2008).
A more detailed description of the method is given in Leahy and Tian (2010).
Basically, careful selection of source and background regions for extraction 
of HI spectra, and correct accounting of radiation transfer effects is needed.
The HI absorption spectra can show absorption features for HI clouds in between
the SNR and the observer, providing a lower limit to distance.
The HI can also show absence of absorption for known HI, such as HI near the tangent point, which gives an upper limit on the distance.

In cases where the SNR is known to be associated with a molecular cloud, the
velocity of the molecular cloud can be measured from CO line observations.
This then provides the kinematic velocity of the SNR.
In other cases the SNR may be associated with an HII region which has a measured
hydrogen recombination line velocity. 
This also provides a kinematic velocity for the SNR.

\section{Results and discussion}

$\,\!$\indent  Fig. 1 shows HI spectra for three different source/background areas
for the SNR G41.1-0.3. 
The top (red) curves are the spectra from the background region.
The middle (black) curves are the spectra from the source region,
and lowest curves in each panel are the difference spectra from the source region. 
The spectra were extracted from the HI data cubes using task 'meanlev'
from the DRAO Export Package Software. 
'meanlev' allows the user to specify source and background regions in the HI data cube by using the radio continuum brightness level
in the 1420MHz continuum image. 
This results in better separation of source and background regions than
other methods which rely on specifying a geometrical shape in the HI
data cube. 
The net result is a better HI absorption spectrum for any given data set.

The current study has yielded a number of kinematic velocities for
SNR in the VGPS survey area. 
For other SNR, we have obtained upper and/or lower limits to kinematic velocities.
The velocities can be converted to distances using a rotation curve model
for the Galaxy. 
We use the recent rotation curve published by Reid et al (2014).
This is based 
mainly on very long baseline interferometry measurements of distances to a number 
of objects in the Galaxy.
The full details of analysis for each SNR in the VGPS area, and comparison with
previously published results, is currently being prepared for publication and release. This includes new (revised or improved) distances to many supernova remnants.

\begin{figure}
\center
\includegraphics[width=\textwidth]{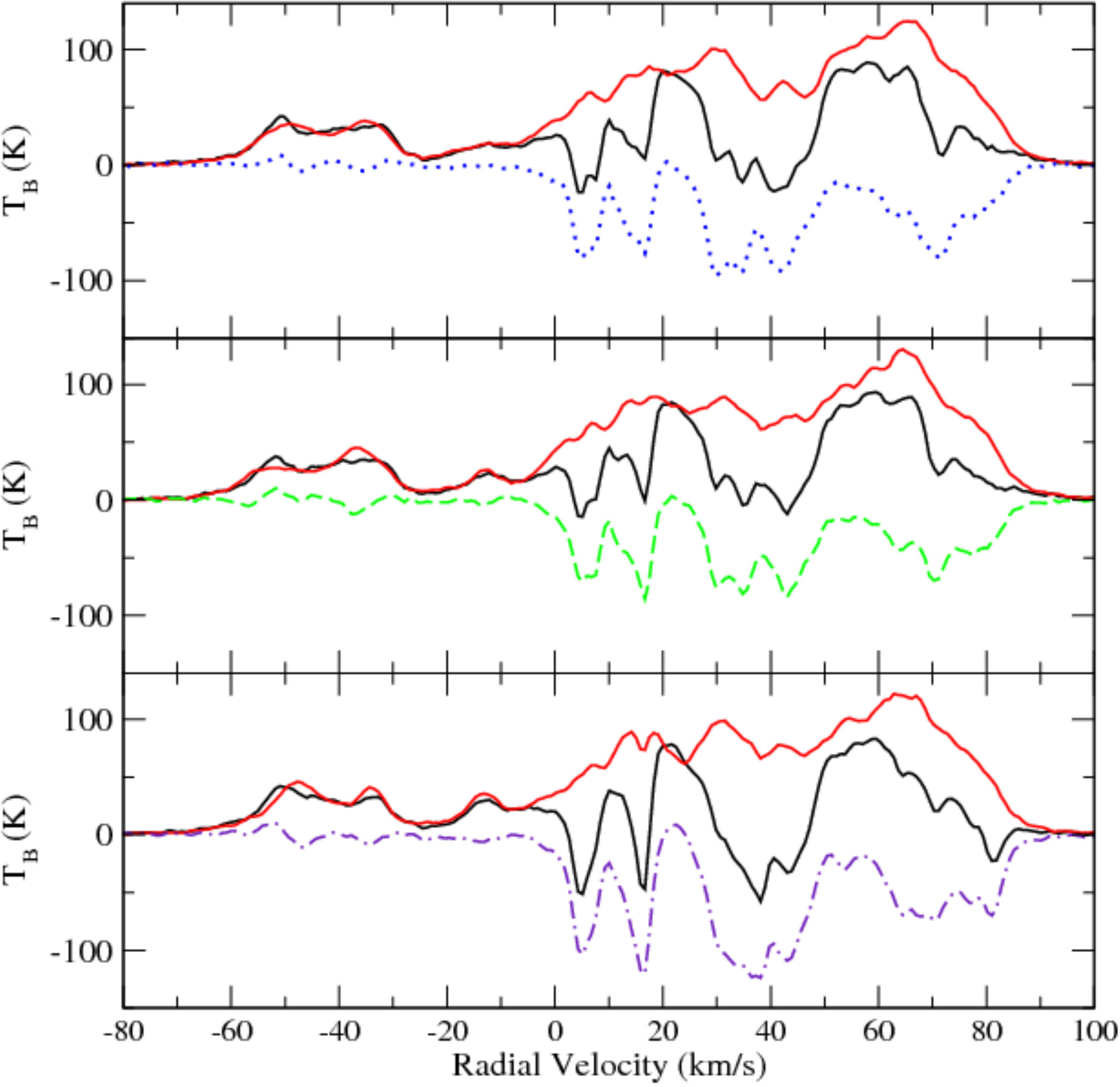} 
\caption{Sample HI spectra for the 3 different source/background regions for the supernova remnant G41.1-0.3 (also known as 3C 397). See text for details.}
\end{figure}

\small  
%
\section*{Acknowledgments}   
%
Support for this work was provided the Natural Sciences and Engineering Research Council of Canada.

\section*{References}
\bibliographystyle{aj}
\small
\bibliography{proceedings}
Stil, J.~M., Taylor, A.~R., Dickey, J.~M., et al.\ 2006, AJ, 132, 1158\\
Tian, W.~W., Leahy, D.~A., \& Wang, Q.~D.\ 2007, A\& A, 474, 541\\
Leahy, D.~A., \& Tian, W.~W.\ 2008, AJ, 135, 167\\
Leahy, D., \& Tian, W.\ 2010, The Dynamic Interstellar Medium: A Celebration of the Canadian Galactic Plane Survey, 438, 365\\
Reid, M.~J., Menten, K.~M., Brunthaler, A., et al.\ 2014, ApJ, 783, 130\\

\end{document}